\documentclass[twocolumn,pre,tightenlines,superscriptaddress]{revtex4-1}

\usepackage{amsmath}
\usepackage{amssymb,amsfonts,latexsym}
\usepackage{bm}
\usepackage[mathcal]{euscript}
\usepackage{graphicx}
\usepackage{epsfig}
\usepackage{color}
\usepackage{xfrac}

\begin{document}

\title{Fore-aft asymmetric flocking}

\author{Qiu-shi Chen}
\affiliation{National Laboratory of Solid State Microstructures and Department of Physics, Nanjing University, Nanjing 210093, China}

\author{Aurelio Patelli}
\affiliation{ISC - CNR, UoS Sapienza, Piazzale Aldo Moro 5, 00185 Roma, Italy}
\affiliation{Service de Physique de l'Etat Condens\'e, CEA, CNRS, Universit\'e Paris-Saclay, CEA-Saclay, 91191 Gif-sur-Yvette, France}

\author{Hugues Chat\'{e}}
\affiliation{Service de Physique de l'Etat Condens\'e, CEA, CNRS, Universit\'e Paris-Saclay, CEA-Saclay, 91191 Gif-sur-Yvette, France}
\affiliation{Beijing Computational Science Research Center, Beijing 100094, China}
\affiliation{Center for Soft Condensed Matter Physics and Interdisciplinary Research, Soochow University, Suzhou 215006, China}

\author{Yu-qiang Ma}
\affiliation{National Laboratory of Solid State Microstructures and Department of Physics, Nanjing University, Nanjing 210093, China}
\affiliation{Center for Soft Condensed Matter Physics and Interdisciplinary Research, Soochow University, Suzhou 215006, China}

\author{Xia-qing Shi}
\affiliation{Center for Soft Condensed Matter Physics and Interdisciplinary Research, Soochow University, Suzhou 215006, China}

\date{\today}

\begin{abstract}
We show that fore-aft asymmetry, a generic feature of living organisms and some active matter systems, can
have a strong influence on the collective properties of even the simplest flocking models. Specifically,
an arbitrarily weak asymmetry favoring front neighbors changes qualitatively the phase diagram of
 the Vicsek model.
 A region where many sharp traveling band solutions coexist is present at low noise strength, {\it below} the
 Toner-Tu liquid, at odds with the phase-separation scenario well describing the usual isotropic model.
 Inside this region, a `banded liquid' phase with algebraic density distribution coexists with band solutions.
Linear stability analysis at the hydrodynamic level suggests that these results are generic and not specific to the Vicsek model.
\end{abstract}

\maketitle

Non-reciprocal (effective) interactions are interesting but rather rare in physical systems \cite{IVLEV}.
They are, however, likely to be more common in active matter.
A nice example of action-reaction symmetry breaking was given
recently by Soto and Golestanian for catalytically active colloids \cite{SOTO-GOLESTANIAN}.
A strong case is that of self-propelled objects interacting solely via volume exclusion: their shape governs their
effective interaction (e.g. aligning or not) and thus their collective behavior \cite{WENSINK}.
In the context of animal and human collective motion, asymmetric interactions are quite generic, and this asymmetry
lies mostly in the relative position and weight of neighbors:
the importance and quality of the information perceived by living organisms usually varies with its origin: In animal groups one often
---but not always, cf. the cannibalistic behavior of locusts in \cite{SIMPSON,BAZAZI}---
expects that frontal stimuli such as neighbor positions matter more to an individual than events taking place in its back.
Somewhat surprisingly, this generic fore-aft asymmetry has not been much investigated {\it per se}.
It is explicitly mentioned in some works \cite{BARBERIS}, and implicitly present in a number of models, see, {\it e.g.} \cite{LUKEMAN}, and
the rather complicated escape-pursuit mechanisms introduced in \cite{ESCAPE-PURSUIT,ESCAPE-PURSUIT2} to describe marching locusts, or the `motion guided attention' of \cite{LEMASSON}.
It can even be found in variants of simple flocking models
such as the Vicsek model, where local alignment of constant-speed particles competes with noise \cite{VICSEK,GREGOIRE,POLAR}.
In \cite{AOV1, AOV2,AOV3,AOV4,AOV5}, the introduction of a limited angle of vision was shown to have an influence on the shape of cohesive moving groups, on the degree of ordering, etc. Asymmetric interactions are also present in `metric-free' models introduced in the context of bird flocks \cite{TOPO1,TOPO2,TOPO3,TOPO4}.

In all cases mentioned above, it was {\it not} shown that fore-aft asymmetry alone can lead to qualitatively new collective phenomena.
Recently, though, the influence of fore-aft asymmetric neighbors was investigated in the context of flocking models
incorporating fast `inertial spin' variables \cite{CAVAGNA,DADHICHI}.
Both these works argue that the combination of asymmetric neighbors and fast variables induces new
collective behavior.

\begin{figure*}[t!]
\includegraphics[width=\textwidth,clip=on]{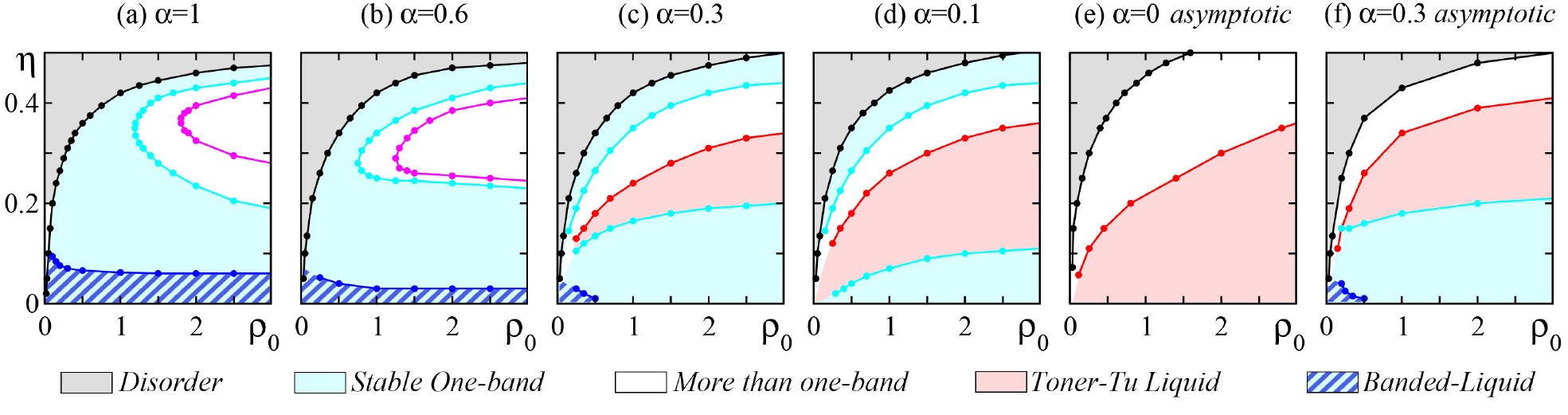}
\caption{
Phase diagram of fore-aft asymmetric Vicsek model with front preference ($\alpha>0$) in $(\rho_0,\eta)$ plane.
Cyan region: stability of the one-band solution.
Red region: Toner-Tu liquid phase.
Blue-stripes region: banded-liquid phase.
Black line: order-disorder transition. In the white regions, one observes solutions with more than one band.
Blue line: upper limit of 'banded liquid' phase.
Cyan line: limit of stability of the one-band solution.
Magenta line: limit of stability of the two-band solution (was only determined in (a,b)).
Red line: upper limit of the Toner-Tu liquid phase.
(a-d) finite-size systems ($L=256$) with $\alpha=1$, $0.6$, $0.3$, and $0.1$.
(e,f) asymptotic ($L\to\infty$) diagrams for $\alpha=0$ (usual Vicsek model, data from \cite{STC}) and $\alpha=0.3$.
}
\label{fig1}
\end{figure*}

In this Letter, we show that fore-aft asymmetry alone has a strong influence on the collective properties of
even the simplest flocking models, devoid of fast inertial variables.
Specifically, an arbitrarily weak asymmetry favoring front neighbors changes qualitatively the phase diagram of
 the Vicsek model, chosen here once more for its archetypical value.
The phase diagram of the usual symmetric Vicsek model ---recalled in Fig.~\ref{fig1}e---
is now well-understood \cite{STC} as the result of a phase separation between a disordered gas and a polarly ordered liquid endowed with the
non-trivial correlations and fluctuations akin to those predicted by Toner and Tu \cite{TT}.
Here we find that the microphase coexistence region is very different even for weak fore-aft asymmetry,
and that the high-order high-density traveling bands characterizing it are much more robust and dominant, forcing one to reconsider
the genericity of the liquid-gas phase-separation scenario.
We also show the emergence, for strong-enough asymmetry, of an ordered phase different from the Toner-Tu liquid,
characterized by the heavy tail of its distribution of local density.
Finally, we derive hydrodynamic equations
which we show to have a linear stability phase diagram in qualitative agreement with our findings at the microscopic level, which
suggests that our results are generic and not specific to the Vicsek model.

Fore-aft asymmetry can be implemented in different ways.
We do so via a minimal modification of the two-dimensional Vicsek model
where the only new feature is that front and back neighbors have respective weights $\frac{1+\alpha}{2}$ and $\frac{1-\alpha}{2}$, so that when $\alpha=1$ (resp. $-1$) only front (resp. back)
neighbors are taken into account (with equal weights).
We thus consider $N$ point particles moving at constant speed $v_0$ in a periodic square domain of linear size $L$.
At each discrete time step $\Delta t=1$, the headings
$\theta$ and positions {\bf r} of all particles are updated in parallel according to
\begin{align}
  \label{eq:model1}
  \theta_j^{t+1}&={\rm arg} [ \langle  \omega_{jk}^t \exp{i\theta_k^t} \rangle_{k\in\mathcal{N}_j}] +\eta\, \xi_j^t \\
  \label{eq:model2}
  {\bf r}_i^{t+1}&= {\bf  r}_i^t+v_0 {\bf e}_i^{t+1}
\end{align}
where $\mathcal{N}_i$ is the unit disk around particle $i$,
$\xi_i^t$ a random angle drawn uniformly in
$[-\pi,\pi]$, $\eta$ sets the noise intensity, ${\bf e}_i^{t+1}$ is the unit
vector pointing in direction given by $\theta_i^{t+1}$ and the weights
$\omega_{jk}^t=\frac{1}{2}(1+\alpha \,{\rm sign}[({\bf r}_k^t-{\bf r}_j^t)\cdot {\bf e}_j^t])$.

\begin{figure}[t!]
\includegraphics[width=\columnwidth]{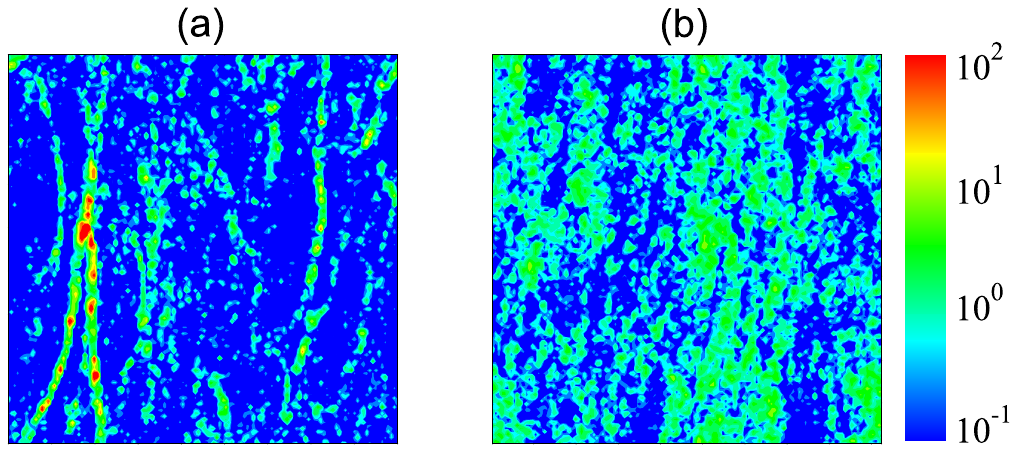}
\includegraphics[width=\columnwidth]{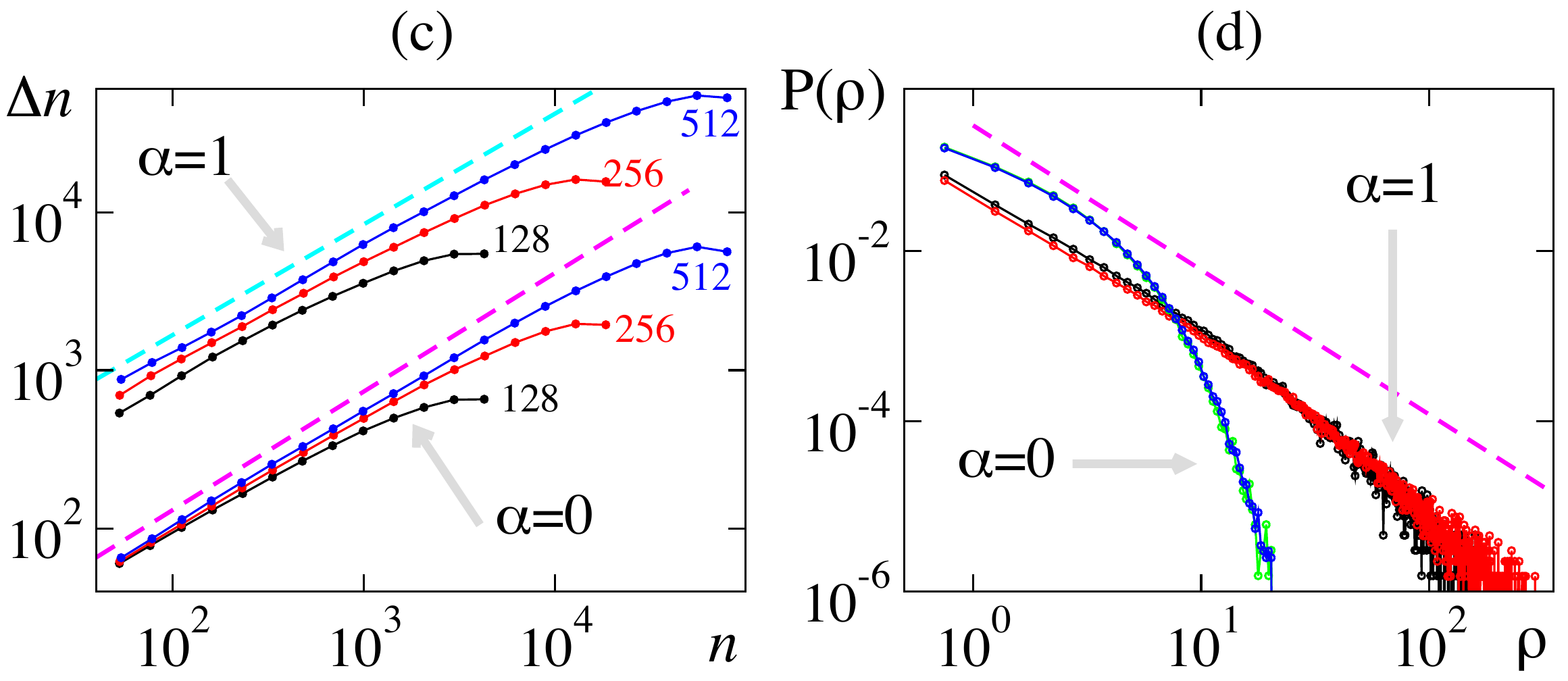}
\caption{
Banded liquid phase observed for $\alpha=1$, $\rho_0=1$, $\eta=0.055$.
(a,b) snapshots of density field calculated over square boxes of linear size $\ell=4$ in a system of size $L=512$ (logarithmic color scale, (a): pure front model $\alpha=1$; (b): Toner-Tu liquid of symmetric Vicsek model $\alpha=0$).
(c) giant number fluctuations: root mean square $\Delta n$ of number of particles $n$ contained in boxes of various linear sizes. Top 3 curves (shifted up for clarity): pure front model ($\alpha=1$). Bottom 3 curves: symmetric Vicsek model ($\alpha=0$). Dashed cyan (resp. magenta) line is a powerlaw of exponent 0.7 (resp. 0.75). The system size is indicated near each curve.
(d) probability distribution function of coarse-grained density
(black and red lines: $\alpha=1$, $L=256,512$; green and blue lines: $\alpha=0$, $L=256,512$). Dashed magenta line is a powerlaw of slope $-1.7$.
}
\label{fig2}
\end{figure}

We first present results obtained in a numerical study
of the model at $\alpha=1$ (only front neighbors are taken into account), varying the global density
$\rho_0=N/L^2$ and the noise intensity $\eta$. The phase diagram at fixed size $L=256$ is presented in Fig.~\ref{fig1}a
(see numerical protocol in Appendix~\ref{AppendixA}).
The order disorder transition line (black line) is similar to but slightly lower than that of the symmetric model, whose phase diagram
is shown in Fig.~\ref{fig1}e.
In strong contrast to this last case, there is no region of homogeneous Toner-Tu liquid at low noise strength.
Instead, traveling band solutions, characteristic
of the coexistence phase, are observed {\it everywhere} below the order-disorder line. More specifically, the solution containing a single band
is stable in the cyan region between the order-disorder line and the C-shaped cyan line in Fig.~\ref{fig1}a. Similarly the 2-band solution is stable from almost the order-disorder line until the C-shaped magenta line
\footnote{There is thus a large region of parameter space where both the one-band and the two-band solutions can be observed depending on initial conditions}.
In the low noise blue-stripes region below the blue line, a phase that we call `banded liquid' (see below and Fig.~\ref{fig2})
coexists with many stable band solutions.
Exploring very large values of $\rho_0$,
we find that the lower branch of the C-shaped curves, and the line delimiting the banded liquid
become largely independent of $\rho_0$. Above the C-shaped curves, at densities $\rho_0\gtrsim 20$, we observed what could be the usual Toner-Tu liquid (not shown), even though it is hard to reach solid conclusions for such parameters.

\begin{figure}[t!]
\includegraphics[width=\columnwidth]{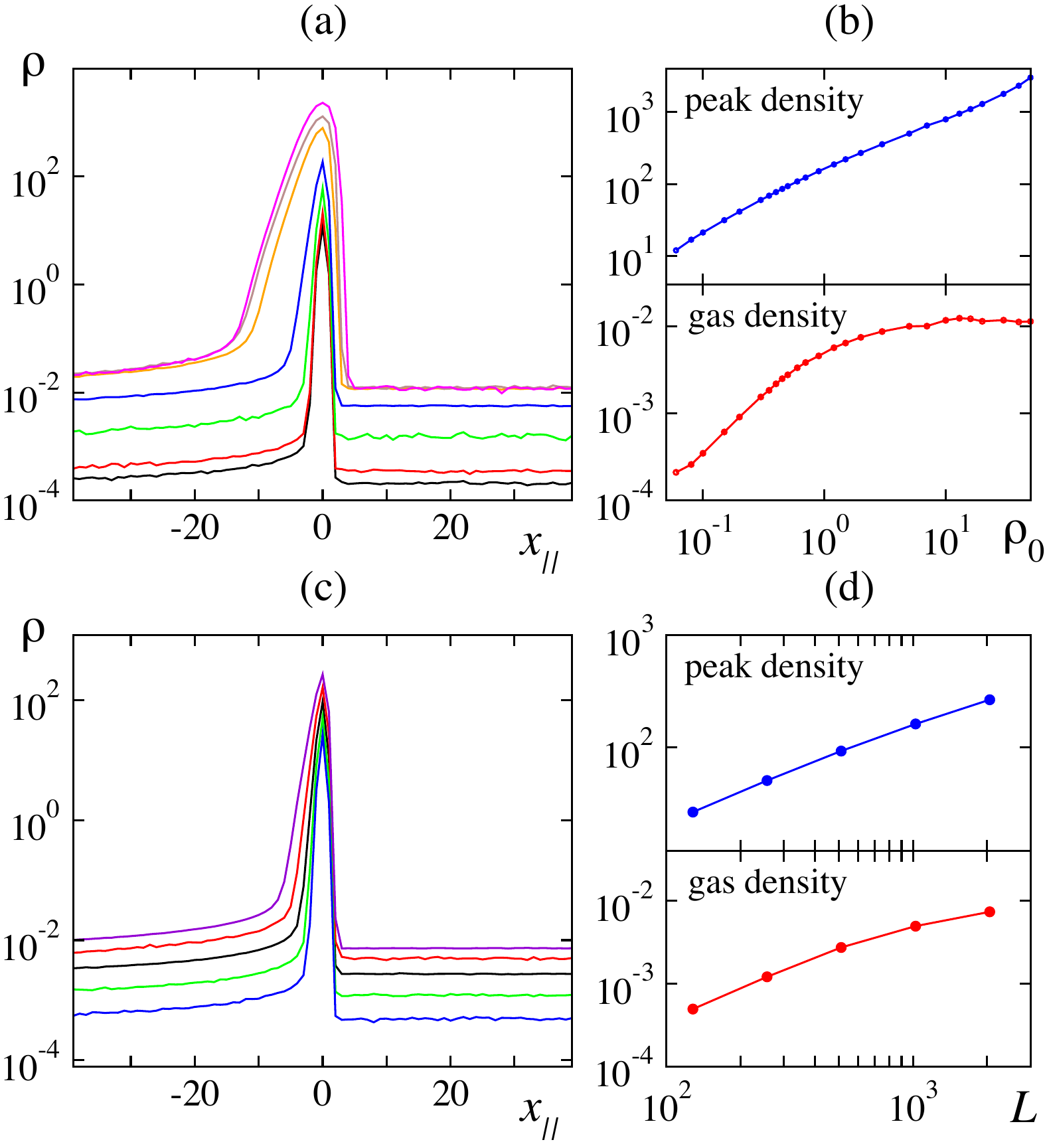}
\caption{
Robustness of the one-band solution ($\alpha=1$, $\eta=0.1$).
(a) time-averaged density profiles at various $\rho_0$ values. From bottom to top: $\rho_0=0.06$, 0.1, 0.3, 1.2, 10, 20, 50 ($L=256$).
(b) peak density (top) and gas density (bottom) vs $\rho_0$ ($L=256$).
(c) time-averaged profiles for $L=128, 256,512,1024,2048$ from bottom upward ($\rho_0=0.25$).
(d) same as (b) but for profiles of (c).
}
\label{fig3}
\end{figure}

One striking feature in the above observations is the robustness of the band solutions, and their extremely sharp fronts and small
width in the low noise region. For $\eta=0.1$ for instance, the one-band solution seems stable for arbitrarily large $\rho_0$,
reaching ever-higher peak densities as $\rho_0$ increases (Fig.~\ref{fig3}a,b).
This solution is also stable for arbitrary large system size $L$, all parameters being fixed (Fig.~\ref{fig3}c,d).
These observations invalidate in practice the liquid-gas scenario (which implies that the number of bands scales linearly with $\rho_0$ and/or $L$ \cite{STC}).
Nevertheless, we are not in presence of some ``condensation'' phenomenon: two bands at short distance from each other do experience mutual repulsion leading them to be asymptotically equidistant (not shown).

We now turn to a description of the `banded liquid'.
Like the ordered liquid of the symmetric model, this phase shows true long-range polar order (not shown) and exhibits
`giant' anomalous number fluctuations with approximately the same scaling exponent (Fig.~\ref{fig2}c).
But unlike the Toner-Tu liquid, it always coexists with band solutions \footnote{It is only reached from certain classes of initial conditions,
for instance initial configurations with random locations but ordered orientation.}.
Moreover, as seen from Fig.~\ref{fig2}a, extremely high density packets
are present, and the band-like structures to which they belong are only seen if some logarithmic density scale is adopted.
This is in contrast with the usual Toner-Tu liquid phase, a snapshot of which is shown in Fig.~\ref{fig2}b for comparison.
Looking at the distribution function of (coarse-grained) density,
a qualitative difference between the two ordered liquids appears:
the banded liquid exhibits a power-law tail with exponent $\sim 1.7$, whereas only an approximately exponential tail is seen for the Toner-Tu liquid (Fig.~\ref{fig2}d).

The results presented so far show that restricting neighbors to those strictly in front ($\alpha=1$) induces strong
differences with the usual ($\alpha=0$) Vicsek model. This is not due to some singularity of the $\alpha=1$ case:
decreasing $\alpha$ from $1$ to $0$, the phase diagram changes smoothly (Fig.~\ref{fig1}(a-d)):
the banded liquid region (blue stripes) shrinks to some triangle near the origin of the $(\rho_0,\eta)$ plane and vanishes completely for $\alpha \lesssim 0.25$.
The C-shaped curves delimiting the stability of the one-band and 2-band solutions move towards the origin and open up into two separate
regions for $\alpha \lesssim 0.4$, while the Toner-Tu liquid appears in a central red region.
Approaching $\alpha=0$, the lower region of stability
of the one-band solution recedes to larger and larger densities,
leaving the familiar phase diagram of the symmetric Vicsek model (Fig.~\ref{fig1}e).

We now present evidence that our main findings hold in the infinite-size limit.
For that numerically demanding task, we focused on the $\alpha=0.3$  case (see Appendix~\ref{AppendixA} for details).
Crossing the blue lines from below in Fig.~\ref{fig1}, the banded liquid disappears in a clearly discontinuous transition,
typically leaving many thin bands.
We located this transition
at fixed $\rho_0$ for systems of increasing size. The transition values $\eta_{\rm bl}(L)$ thus defined
{\it increase} with $L$ and converge to a finite asymptotic value exponentially,
$\eta_{\rm bl}(\infty)-\eta_{\rm bl}(L) \sim \exp(-kL)$, as expected in first-order phase transitions.
Repeating this for different $\rho_0$ values, we find that the banded liquid region
converges to the blue line in Fig.~\ref{fig1}f.
(For larger $\alpha$ values, the asymptotic banded liquid region remains unbounded in $\rho_0$.)

Similarly, we tracked the stability region of the one-band solution (again, some details are found in Appendix~\ref{AppendixA}). We found that its lower part converges to a well-defined domain delimited by the cyan line in Fig.~\ref{fig1}f.
The upper part of the stability region of the one-band solution behaves like in the ($\alpha=0$) Vicsek model: it quickly converges onto the
order-disorder line as $L$ increases (and is thus undistinguishable from the black order-disorder line in Fig.~\ref{fig1}f).
Like for the symmetric model, there exists nevertheless a
coexistence region containing smectic arrangements of band solutions (white area in Fig.~\ref{fig1}f).
We thus conclude that the asymptotic phase diagram of our model for $\alpha=0.3$ is qualitatively different from that
of the symmetric model (compare Figs.~\ref{fig1}e and \ref{fig1}f). It comprises an extra `banded-liquid' phase as well as a
second region supporting band solutions that seems unbounded to the right.

In fact, the phase diagram seems to comprise a lower band region for {\it any}
finite $\alpha$ value: the one-band solution can be found stable for values as low as $\alpha=0.01$,
provided that $\rho_0$ is large enough. For instance, working at fixed low noise $\eta=0.03$ and $L=256$,
we find that the one-band solution is stable for at least millions of timesteps for $\rho_0\ge \rho^*(\alpha)$,
with $\rho^*\simeq 0.65, 1.1, 1.8, 3$ for $\alpha=0.06, 0.04, 0.02 ,0.01$ and that $\rho_0^*$ {\it decreases}
with increasing system size (at fixed $\alpha$).
This suggests nothing less than the singularity and `fragility' of the classic Vicsek model ---and thus, presumably,
of many flocking models in its class--- with respect to fore-aft asymmetry.

We finally report evidence of the robustness of our findings. First, they are not specific to the variant of the Vicsek model defined by
Eqs.~(\ref{eq:model1},\ref{eq:model2}): implementing instead a restricted angle of vision $[\theta_i-\delta;\theta_i+\delta]$ around
the particle's headings $\theta_i$, we found phase diagrams similar to those of Fig.~\ref{fig1} for $\delta\in[\frac{\pi}{2},\pi]$ (not shown).
Next, we now show briefly how one can account for our results at the hydrodynamic level (details can be found in Appendix~\ref{AppendixB}).
We use a variant of the `Boltzmann-Ginzburg-Landau' method used successfully for Vicsek-style models \cite{BGL}.
This approach, or for that matter any one based on a controlled truncation and closure of a kinetic equation
governing the one-body distribution function  $f(\theta,\vec{r},t)$
can only produce the following Toner-Tu equations
coupling the density and the ordering fields $\rho$ and ${\bf w=\rho\,{\bf P}}$ (where ${\bf P}$ is the polar order field):
\begin{eqnarray}
\partial_t \rho + \nabla \!\cdot\! {\bf w} &=& D \Delta {\bf w} \label{hydro1} \\
\partial_t {\bf w} + \frac{1}{2}\nabla\rho&=&  [\mu \!-\! \xi |{\bf w}|^2]{\bf w}+D_{\rm i} \Delta{\bf w} +D_{\rm a} \nabla(\nabla\!\cdot\!{\bf w}) \nonumber \\
&&\!\!\!\!\!\!-\lambda_1({\bf w} \!\cdot\! \nabla) {\bf w} -\lambda_2 (\nabla\!\cdot\! {\bf w}) {\bf w} - \lambda_3 \nabla |{\bf w}|^2
\label{hydro2}
\end{eqnarray}
but the explicit form of the $\rho$-dependent transport coefficients does depend on the precise kinetic equation used.
Here we write a standard Boltzmann equation with effective positional diffusion
\footnote{The `spurious instability' found at low noise at hydrodynamic level in the symmetric case without positional diffusion was found to
disappear upon the addition of even weak diffusion \cite{BGL}. Here diffusion has the same regularizing effect and we thus add it 'by hand'.}
and the collision integral:
\begin{eqnarray}
	I_{\rm col}[f] &=& \!\!\int \!\! d\theta_1d \theta_2 \,f(\theta_1,\vec{r})f(\theta_2,\vec{r}) \int \!\! d\phi \,K(\theta_2\!-\!\theta_1,\phi) \nonumber \\
	&\,&\times [P_\eta\ast \delta(\Psi(\theta_1,\theta_2))-\delta(\theta_1-\theta)].
	\label{eq:CollisionIntegral}
\end{eqnarray}
where $P_\eta$ is the angular noise distribution,
$\Psi(\theta_1,\theta_2)={\rm arg}[\exp ( \frac{i}{2}(\theta_1+\theta_2))]$ the polar alignment function, and $\ast$ the convolution operator.
The main difference with respect to the traditional modeling of collisions
is encoded in the kernel $K(\theta_2-\theta_1,\phi)$ and the integral over $\phi$, the angular position of particle 2 in the reference frame of particle 1
 (see Fig.~\ref{fig4}a).
Following the usual route,
collisions would be restricted to those with particles ahead {\it approaching} the reference particle
and would hence discard the crucial, persistent, aligning events with front particles going `away' from the focus particle.
Here, on the contrary, the structure of $K$ favors those collisions, something we encode in the compact expression:
\begin{equation}
	\label{eq:CollisionKernel}
	K(\Delta,\phi) =\vert \sin (\Delta/2)\vert
	\left[\Phi(\Delta,\pi,\phi) + \Phi(\pi, \Delta,\phi+\pi) \right]
\end{equation}
where $\Phi(a,b,x)=\Theta(a-b) \Theta(x-\frac{a}{2})\Theta(\frac{b}{2}-x)$ with $\Theta$ the Heaviside step function
\footnote{Note that this choice stands for the extreme case $\alpha=1$. Interpolating between this and the usual symmetric
neighbors case $K(\Delta) = | \sin (\Delta/2)|$ if left for future work.}.

The resulting expressions for the transport coefficients of Eqs.~(\ref{hydro1},\ref{hydro2}) are given by {Eqs.~(\ref{AP_1}-\ref{AP_2}) in 
Appendix~\ref{AppendixB}.
The study of the existence and linear stability of the homogeneous solutions of Eqs.~(\ref{hydro1},\ref{hydro2}) is summarized
in the $(\rho_0,\sigma)$ phase diagram presented in Fig.~\ref{fig4}b, where $\sigma$ is the root-mean-square (rms) of the microscopic noise $P_\eta$.
Like for the microscopic model, the basic order/disorder transition line defined by $\mu=0$ is lower than the one found with symmetric interactions.
Bordering this line, we find the usual longitudinal, `banding' instability of the ordered solution $|{\bf w}|^2=\mu/\xi$,
but also, at low noise, a new, large, transversal
instability region which almost suppresses the longitudinal instability at low densities.
The exchange of instability direction is in fact the result of growing `transverse' lobes in Fourier space (see Appendix~\ref{AppendixB}).
The new transverse instability causes the Toner-Tu liquid to be unstable at low densities, in agreement with the phase
diagram of the microscopic model.
Investigating whether this agreement carries over to the nonlinear and fluctuating level and possibly reveals
the existence of a banded liquid regime for the {\it stochastic} version of hydrodynamic equations (\ref{hydro1},\ref{hydro2})
is a difficult task left for future studies.
Preliminary simulations at the deterministic level are encouraging in this respect:
in the transverse instability in Fig.~\ref{fig4}b, we observe
very sharp band solutions, reminiscent of those observed in the microscopic model.

\begin{figure}[t!]
\includegraphics[width=\columnwidth]{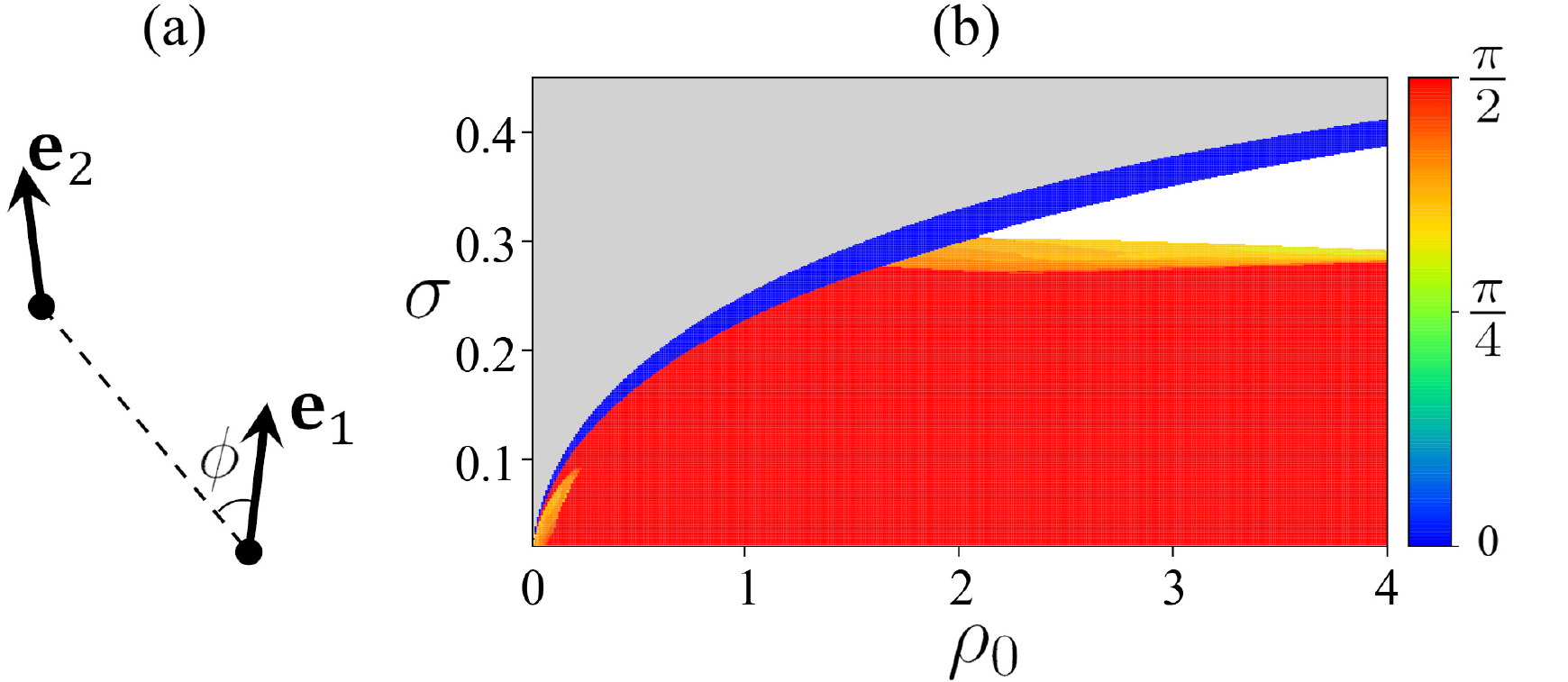}
\caption{
(a) Sketch of the geometry of interaction between particles where particle $1$ is taken as the reference particle.
(b) phase diagram in the $(\rho_0,\sigma)$ plane showing the linear stability analysis of the homogeneous ordered solution
$|{\bf w}|^2=\mu/\xi$ of Eqs.~(\ref{hydro1},\ref{hydro2}).
The color coding indicates the direction of the most unstable mode compared to that of the polar order. In the white region, the homogeneous order is stable; the grey region denotes the stable disordered phase.
}
\label{fig4}
\end{figure}

To summarize, even weak fore-aft asymmetry, a generic feature of living organisms and some active matter systems, can
have a strong qualitative influence on the collective properties of even the simplest flocking models.
Specifically, an arbitrarily small asymmetry favoring front neighbors changes qualitatively the phase diagram of
 the Vicsek model.
A region where many sharp traveling band solutions coexist is present at low noise strength, {\it below} the usual
 Toner-Tu liquid, an observation that forces ones to revisit the phase-separation scenario put forward for symmetric flocking models.
 Inside this region, a `banded liquid' phase with algebraic density distribution coexists band solutions.
Stability analysis at the hydrodynamic level suggests that these results are generic and not specific to the Vicsek model.

Future work will try to connect our results to the interesting predictions made in \cite{CAVAGNA,DADHICHI} for asymmetric
models with fast underdamped inertial spin variables.
More generally, our results indicate that non-reciprocal interactions in active matter deserve further study.

\acknowledgments
This work is supported by the National Natural Science Foundation of China (No. 11635002 to H.C. and X.S.; Nos. 91427302, 11474155 to Y.M.; Nos. 11474210, 11674236 to X.S.).

\appendix
\section{Numerical protocols and details}	
\label{AppendixA}

	{\it Numerical protocol for the phase diagrams in Figs.~1a-d:} in the cyan region, the one-band solution is observed for at least $10^6$ timesteps. Similarly, the two-band solution is observed for at least $10^6$ timesteps left of the magenta line in Figs.~1a,b. The black, cyan, and magenta transition lines obtained are sharply defined because the lifetime of the one-band and two-band solutions quickly and surely becomes very short across them. The blue line marking the limit of the banded liquid phase is again defined by the condition that the banded liquid is observed for at least $10^6$ timesteps. The corresponding transition is sharp, as the breakdown of the banded liquid quickly leads to a solution containing a number of very thin bands, but it has all the features of a first-order phase transition. In particular, the breakdown of the banded liquid follows the {\it nucleation} of some local thin-band(s). Our protocol to locate the transitions points was that, over 100 runs, at least 50 of them led to band solutions within $10^6$ timesteps.
	
	{\it Convergence of transition points at fixed parameters as $L\to\infty$ (Fig.~1f):} to determine each of the transition points shown in the asymptotic phase diagram in Fig.1f, we followed the same protocol as outlined above for different system sizes. This yielded a series of transition values which typically converges exponentially to a finite asymptotic value. An example is shown in Fig.~\ref{figS1}a.
	
\begin{figure}[t!]
\includegraphics[width=\columnwidth]{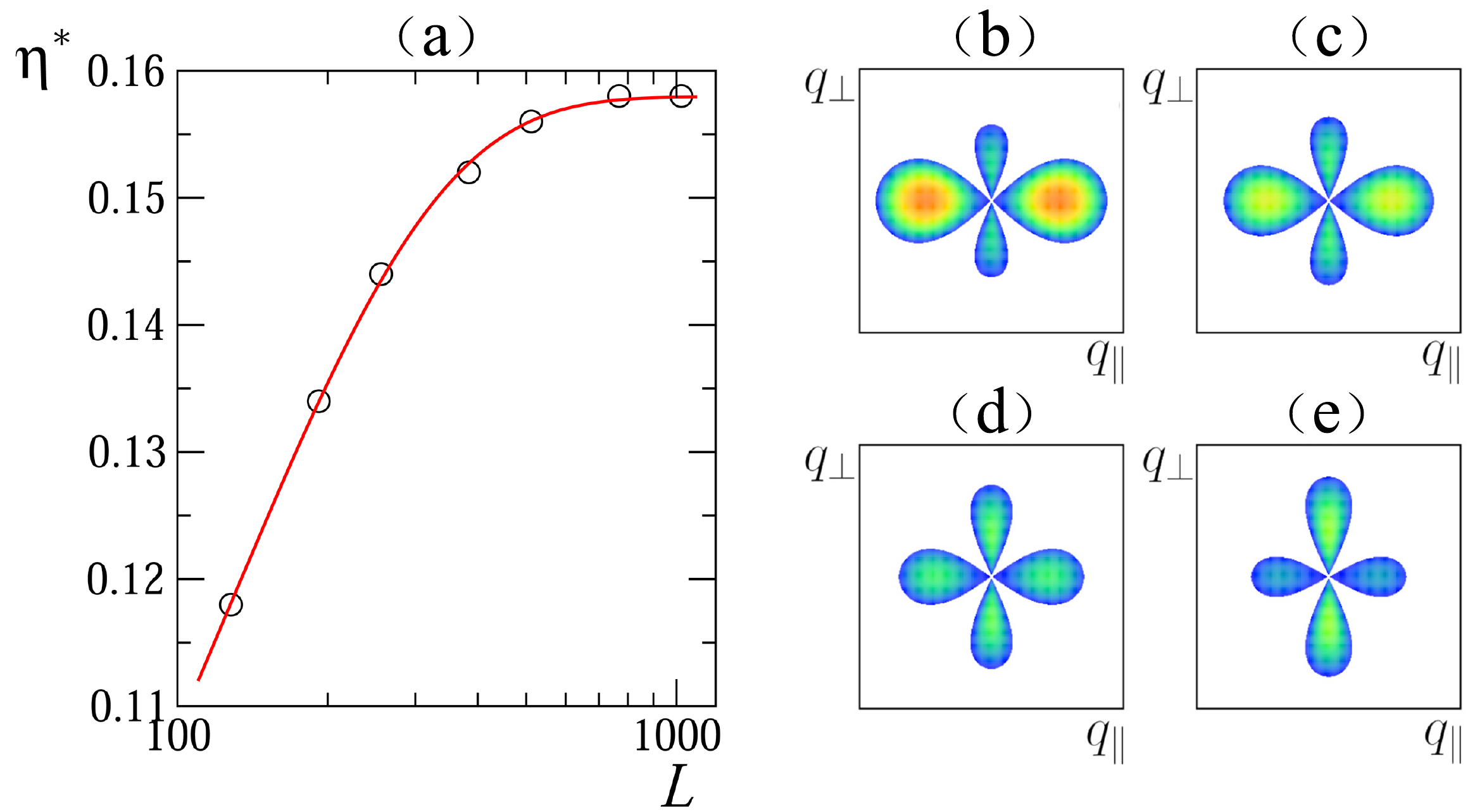}
\caption{(a): Convergence with system size of the transition point $\eta^*(L)$ delimiting the lower part of the stability region of the one-band solution ($\alpha=0.3$, $\rho_0=0.5$). The red solid line is a fit $\eta^*(L)=\eta^*_\infty -a\exp(-bL)$ with $\eta^*_\infty\simeq 0.158$.
(b-e): Unstable modes for the homogeneous solution $|{\bf w}|^2=\mu/\xi$ at $\rho_0=1.5$ in the $(q_\|,q_\perp)$ plane
($q_\|, q_\perp \in [-0.025,0.025]$). Color codes for the growth rate only if it is positive, from blue$=0$ to red$=10^{-3}$.
((b) $\sigma = 0.267$, (c) $\sigma = 0.265$, (d) $\sigma = 0.263$, (e) $\sigma = 0.261$).
}
\label{figS1}
\end{figure}

\section{Derivation of hydrodynamic equations and linear stability analysis}
\label{AppendixB}

	{\it Derivation of hydrodynamic equations:}
	The Boltzmann kinetic equation describing the evolution of the single particle distribution function $f=f({\bf r},\theta,t)$ reads:
	\begin{equation}
	\partial_t f + v_0{\bf e}(\theta)\cdot\vec{\nabla} f = D\triangle f + I_{\rm dif}[f] + I_{\rm col}[f]
	\end{equation}
	where ${\bf e}(\theta)$ is the unit vector along $\theta$ and $I_{\rm dif}[f]=-f+(P_\eta \ast f)$ is the angular self-diffusion integral, with $P$ is the distribution of the
	angular noise and $\ast$ stands for the convolution operator.
	In addition to this rotational diffusion, we consider also spatial diffusion expressed by the term $D\triangle f$. Our Vicsek-style model
	does not have explicit positional diffusion, but its discrete-time dynamics generates it. Here, we introduce it explicitly mostly because
	this term, with $D$ large enough, removes the spurious instability of the homogeneous state found away from the order/disorder line in the resulting hydrodynamic equations \cite{BGL}.
	
	For the standard (isotropic) Vicsek model, the collision between two particles can happen along any approaching direction.
	For the 'front-biased' model ($\alpha>0$), in contrast, we favor the `dominant' aligning interactions, which are those taking place
	effectively with particles already following the reference particle.
	These collisions are frequent, the relative velocity of the two particles is small,
	when the relative direction of particle 2 lies in $\phi\in[\pi/2,3\pi/2]$. This is encoded in the
	$K(\theta_2-\theta_1,\phi)$ given in the main text.
	
	The hydrodynamic equations are then obtained following the usual procedure: expanding the Boltzmann equation in Fourier series of
	the angular variable $\theta$, ($f({\bf r},\theta,t)=\sum_{-\infty}^{\infty} f_k({\bf r},t) \exp (ik\theta)$),
	leading to a hierarchy of partial differential equations governing the fields $f_k$. Using a propagative scaling ansatz \cite{BGL}, this hierarchy
	is truncated and closed, leading to the usual `Toner-Tu' equations given in the text with the following transport coefficients:
	\begin{eqnarray}
	\mu(\rho) &=& P_1 - 1 + \frac{1}{12\pi} \left(8P_1 - (16 - 3\pi) \right) \rho \label{AP_1}\\
	\xi &=& -\frac{8P_1 + 15\pi - 48}{120\pi^2\mu_2} \left[\frac{4}{3} - \frac{\pi}{2} + (4 - \pi) P_2 \right]\\
	D_i &=& D - \frac{1}{4\mu_2} \;\;\;\;	D_a = 0 \\
	\lambda_1 &=& -\kappa_2-\kappa_1 \;\;\;\;  \lambda_2 = -\kappa_2 + \kappa_1\;\;\;\; 	\lambda_3 = - \frac{\lambda_2}{2}
	\end{eqnarray}
	with
	\begin{eqnarray}
	\mu_2 &=& P_2 - 1 - \frac{1}{2\pi}  \left[ \left(\frac{8}{3} - \pi \right)P_2 + \left( \frac{56}{15}-\pi\right)\right]\rho_0\\
	\kappa_1 &=& \frac{1}{2\pi\mu_2}\left(\frac{4}{3} - \frac{\pi}{2} + (4 - \pi) P_2 \right) \\
	\kappa_2 &=& -\frac{1}{120\pi\mu_2}\left(8P_1 + 15\pi - 48 \right) \label{AP_2}
	\end{eqnarray}
	where $P_k=\int_{-\infty}^\infty d\sigma P_\eta(\sigma) \exp(ik\sigma)$.

	{\it Linear stability analysis of the ordered solution:}
	The homogeneous ordered solution $|{\bf w}|^2=\mu/\xi$ to the hydrodynamic equations is unstable in a large region. Near the order-disorder transition line, the most unstable modes are longitudinal. Decreasing $\sigma$, the most unstable modes become transversal ones.
	In Figs.~\ref{figS1}b-e, we show how this happens in the $(q_\|,q_\perp)$ Fourier space of perturbations. Only modes with positive growth rate are shown. They form 4 `lobes' whose relative strength varies with $\sigma$.

\end{document}